\documentclass[
reprint,
amsmath,amssymb,aps,
]{revtex4-2}

\usepackage{graphicx}
\usepackage{dcolumn}
\usepackage{bm}
\usepackage[mathlines]{lineno}
\renewcommand{\figurename}{Fig}

\begin{document}


\title{Magnetic-field-induced insulator--metal transition in W-doped VO$_2$ at 500~T}

\author{Yasuhiro H. Matsuda}
\email{ymatsuda@issp.u-tokyo.ac.jp}
\author{Daisuke Nakamura} 
\author{Akihiko Ikeda} 
\author{Shojiro Takeyama} 
\affiliation{%
The Institute for Solid State Physics, The University of Tokyo,5-1-5 Kashiwa, Chiba 277-8581, Japan
}%

\author{Yuji Muraoka}
\author{Yuki Suga}
\affiliation{
 Research Institute for Interdisciplinary Science, Okayama University, 3-1-1 Tsushima-naka, Tsushima, Kita-ku, Okayama 700-8530, Japan
}%

\date{\today}
\maketitle

\textbf{Metal--insulator (MI) transitions in correlated electron systems have long been a central and controversial issue in material science. 
Vanadium dioxide (VO$_2$) exhibits a first-order MI transition at 340 K \cite{Morin1959}. 
For more than half a century, it has been debated whether electronic correlation \cite{Zylbersztejn1975, Rice1994,Huffman2017} 
or the structural instability due to dimerised V ions \cite{Goodenough1960, Wentzcovitch1994, Eyert2002, Hiroi2015} is the more essential driving force behind this MI transition.   
Here, we show that an ultrahigh magnetic field of 500~T renders the insulator phase of tungsten (W)-doped VO$_2$ metallic. 
The spin Zeeman effect on the $d$ electrons of the V ions dissociates the dimers in the insulating phase, resulting in the delocalisation of electrons. 
Because the Mott--Hubbard gap essentially does not depend on the spin degree of freedom, 
the structural instability is likely to be the more essential driving force behind the MI transition. }
\\





The Mott--Hubbard insulator is a class of materials in which strong electron correlation disturbs the motion of electrons and electrons are localised \cite{Imada1998}. 
The magnetic ground state of many Mott-Hubbard insulators shows antiferromagnetic order, and high-$T_c$ superconductivity occurs near 
the magnetic-order phase, indicating that the spin degree of freedom is important to understand their peculiar electronic states.  
On the other hand, several insulators possess strong electron correlation, but their magnetic ground state is a spin singlet (nonmagnetic) \cite{Imada1998}. 
VO$_2$ is one such material. 
Strong electron correlation has been claimed to be necessary for understanding the large energy gap of 0.7~eV in the low-temperature 
insulating phase of VO$_2$ \cite{Zylbersztejn1975, Rice1994, Huffman2017}. 
A key feature of the MI transition of VO$_2$ is that it occurs along with a structural transition from a high-temperature rutile (tetragonal) phase 
to a low-temperature monoclinic phase \cite{Morin1959, Eyert2002, Hiroi2015}.
Vanadium dimers are formed in the low-temperature monoclinic phase, and $d$ electrons of the V ions (V$^{4+}$: $d^1$) form metal--metal bonding with a molecular orbital. 
The $d$ electrons are localised in the molecular orbital of the V--V dimer, which can result in the insulating nature \cite{Goodenough1960, Wentzcovitch1994, Eyert2002, Hiroi2015}. 


Many theoretical and experimental studies have been conducted to determine whether the structural instability due to dimerisation \cite{Goodenough1960, Wentzcovitch1994, Eyert2002, Hiroi2015} or electron 
correlation (Mott physics) \cite{Zylbersztejn1975, Rice1994,Biermann2005,Huffman2017,Najera2017} is the more essential driving force behind the 
MI transition.  
Understanding of the microscopic mechanism of the MI transition of VO$_2$ is also important for its practical applications as sensors 
and switching devices; abrupt changes in resistivity and optical absorption at the transition temperature 340~K are useful for them \cite{Jia2018}.  
The manipulation of electron spins by a magnetic field can shed light on the problem of MI transition because the dimerisation can be suppressed by the Zeeman energy in magnetic fields. 
One may envision that the molecular orbital between the V atoms collapses by the forced alignment of the spin direction because the bonding orbital is formed only with 
two electrons having opposite spin states. 
The advantage of utilising a magnetic field is that the electronic state can be modified through the Zeeman splitting while maintaining the quantum-mechanical electron correlation 
that is significant at low temperatures. 
Since the Mott--Hubbard gap is expected to be insensitive to the spin state, the magnetic field cannot affect the insulating nature if the Mott physics is the more essential 
driving force behind the MI transition. 
Ultrahigh magnetic fields with Zeeman energy at least comparable to the thermal energy at the MI transition temperature ($T_{\textrm{MI}}$) would be required 
to investigate the potential magnetic-field-induced metallisation of VO$_2$.

In the present study, we experimentally demonstrate that the insulating dimerised state can be transformed to a metallic state by a strong magnetic field 
of 500~T in W-doped VO$_2$, the MI transition temperature of which is controlled to approximately 100~K \cite{Shibuya2010}. 
We performed magneto-transmission experiments using a near-infrared laser line and found a significant decrease in the transmitted light intensity 
at the ultrahigh-magnetic-field region, which is distinct evidence of the field-induced insulator--metal (IM) transition. 
The observed onset field of the transition at 14~K is approximately 120~T, and its Zeeman energy corresponds to 162~K when $g$ = 2 and $S$ = 1/2, 
where $g$ and $S$ are the $g$-factor and spin quantum number, respectively. 
The observed magnetic-field-induced metallisation indicates that the dimerisation is a more essential driving force than the electron correlation for the MI transition 
in VO$_2$.



\begin{figure}
\begin{center}
\includegraphics[width=7cm]{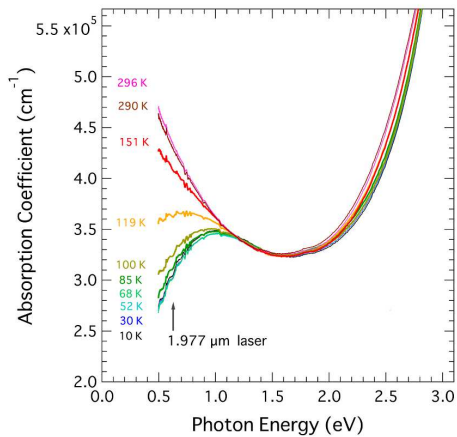}
\caption{\textbf{Optical absorption spectra at different temperatures without external magnetic fields.} The arrow indicates the 
energy of the laser used for the magneto-transmission experiments.}
\label{zero_spectra}
\end{center}
\end{figure}

Figure~\ref{zero_spectra} shows optical absorption spectra of a V$_{1-x}$W$_x$O$_2$ ($x$ = 0.06) thin film at different temperatures without external magnetic fields. 
The broad absorption peak around 1~eV can be attributed to the excitation from the bonding $d_{||}$ orbital to the non-bonding 
$\pi^*$ orbital of the $d$ electrons of V$^{4+}$ ($d^1$) in the octahedron of oxygen ions \cite{Gavini1972,Eyert2002,Okazaki2006,He2016}. 
The $d_{||}$ and $\pi^*$ orbitals originate from the $t_{2g}$ state in the crystal field with cubic symmetry. 
The strong increase in absorption with photon energy increasing beyond approximately 1.8~eV is attributed to the 
transition of charge transfer from the vanadium 3$d$-like band to the oxygen 2$p$-like band \cite{Okazaki2006}.

The absorption below approximately 1.2~eV is found to increase with decreasing photon energy at a high temperature of, e.g., 296~K. 
The absorption becomes less significant as the temperature decreases. 
This behaviour is accounted for by the opening of the energy gap and a change in the number of conduction electrons \cite{Okazaki2006, Lee2012}. 
The significant decrease of absorption with decreasing temperature at energies such as 0.627~eV (corresponding to the laser line wavelength of 1.977~$\mu$m 
used for the magneto-transmission experiment) directly reflects the MI transition. 
The observed temperature dependence of the spectra is very similar to the previously reported result for V$_{1-x}$W$_x$O$_2$ ($x =0.05$) \cite{Lee2012}. 

\begin{figure}
\begin{center}
\includegraphics[width=6.5cm]{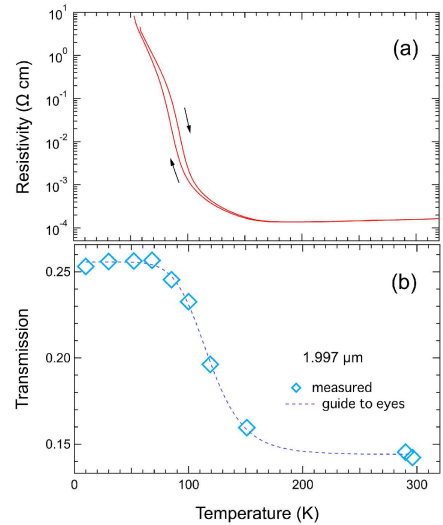}
\caption{\textbf{Temperature dependence of electrical resistivity and optical transmission of a V$_{1-x}$W$_x$O$_2$ ($x =0.06$) thin film.} \textbf{a}, Temperature dependence of electrical resistivity ($\rho$). The arrows indicate heating and cooling processes. 
\textbf{b}, Temperature dependence of the optical transmission at 1.977~$\mu$m.  The error bar is as large as the size of the diamond mark in the plot. 
}
\label{rho_trans}
\end{center}
\end{figure}

Figures~\ref{rho_trans} (a) and (b) show, respectively, the temperature dependence of electrical resistivity ($\rho$) and that of the optical transmission at 1.977 $\mu$m of three layers of a 
V$_{1-x}$W$_x$O$_2$ ($x =0.06$) thin film with a total thickness of 45 nm. 
The MI transition occurs around 100~K, and 
the transmission below approximately 70~K is nearly independent of temperature. 
The curve-fitting analysis of the spectra, taking into account the DC electrical resistivity, shows that the effect of closing of the energy gap 
(change in the absorption band) is more significant than that of the free-carrier absorption. 
When the carrier density is less than approximately 10$^{25}$ m$^{-3}$, the transmission at 0.627~eV is nearly independent of the carrier density. 
Details of the fitting are described in the Supplementary Information. 
The hysteresis observed in the temperature dependence of $\rho$ indicates the first-order nature of the MI transition, 
although it is not observed very clearly in the temperature dependence of the transmission. 
The transmission is measured with both cooling and heating processes, and Fig.~\ref{rho_trans} (b) plots the averaged results. 

\begin{figure}
\begin{center}
\includegraphics[width=8cm]{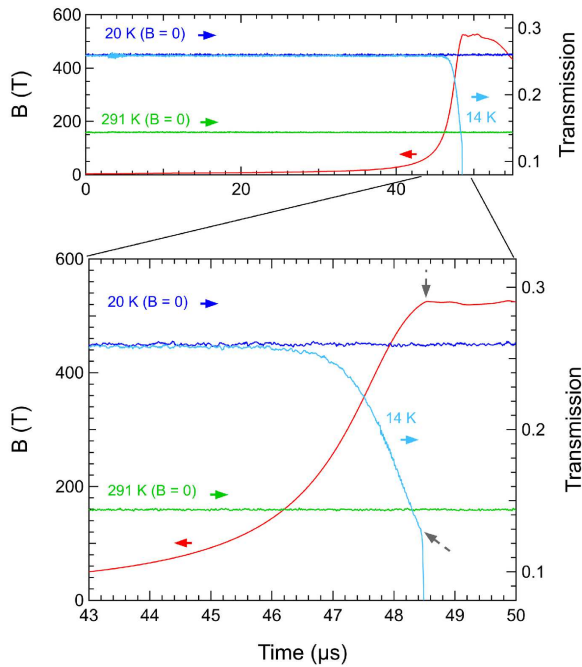}
\caption{\textbf{Temporal evolutions of the magnetic field and optical transmissions.} Upper panel: evolutions of the magnetic field (red curve) and the optical transmission at 14~K (light-blue curve). 
The temporal evolutions of the transmission at 20~K and 291~K before applying the magnetic field are also shown with blue and light-green curves, respectively. 
Lower panel: magnified view of the upper panel. 
The two dashed arrows show the breaking points of the measurements of the magnetic field and optical transmission, respectively. 
They are found to occur at nearly the same time as the timing of termination of the magnetic-flux compression process \cite{Nakamura2018}. 
}
\label{time_chart}
\end{center}
\end{figure}

Ultrahigh magnetic fields ($B$) of up to 520~T are applied perpendicular to the thin-film plane ($B || c$, where $c$ is the crystal axis of the rutile structure), 
and the optical transmission at 1.977~$\mu$m is simultaneously measured. 
As shown in the upper panel of Fig.~\ref{time_chart}, the magnetic field (red curve) increases with time ($t$) and reaches 520~T at 42.8~$\mu$s 
after the ignition of gap switches of the capacitor bank power supply at $t=0$. 
Figure~\ref{time_chart} shows the $t$ evolution of the transmitted light intensity at 14~K with a light-blue curve. 
The lower panel of the figure shows a magnified view of the region from 50 to 520~T. 
%
The optical transmission at 20 and 291~K without magnetic fields is measured immediately before conducting the destructive ultrahigh-magnetic-field experiment, 
clearly showing the transmission levels in the insulating and metallic phases, respectively. 

The transmission at 14~K under a magnetic field starts to show a gradual decrease at approximately 45~$\mu$s when the magnetic field reaches approximately 100~T;  
subsequently, the decrease is accelerated with increasing magnetic field. 
Then eventually, the transmission under fields greater than 500~T is less than that in the high-temperature metallic phase (291~K), which 
clearly shows that the insulating phase of V$_{1-x}$W$_x$O$_2$ ($x =0.06$) is transformed to a metallic phase under ultrahigh magnetic fields exceeding 500~T. 
Because the transmission at 14~K keeps decreasing beyond the value corresponding to the high-temperature metallic phase, the electrical resistivity is expected to be 
lower in the high-field metallic phase than in the high-temperature metallic phase. 
A lower scattering rate of electrons is expected owing to reductions in phonon and magnetic scattering in the novel low-temperature high-magnetic-field metallic phase. 

\begin{figure}
\begin{center}
\includegraphics[width=8.5cm]{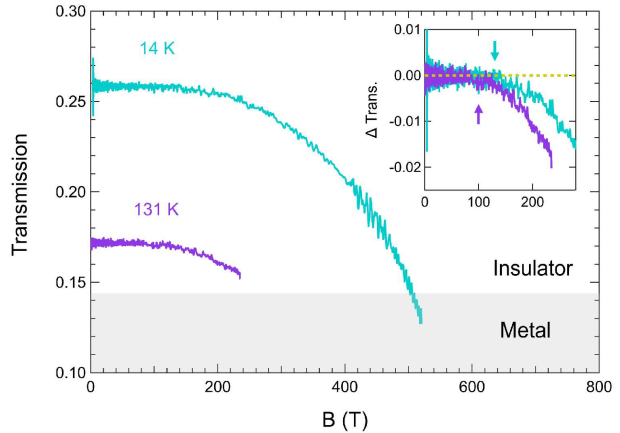}
\caption{\textbf{Magnetic-field dependence of the optical transmission at 14 and 131~K.}  
The transmission is plotted before the breaking point of the destructive measurement. 
The shaded area indicates transmission lower than that in the high-temperature metallic phase at 291~K. 
Inset: transmission change $\Delta \rm{Trans.}$ as a function of the magnetic field. 
The dashed yellow line corresponds to $\Delta \rm{Trans.}=0$. 
The light-blue and purple arrows indicate the field positions ($B^*$) where $\Delta \rm{Trans.}$ deviates from zero. 
}
\label{MagT}
\end{center}
\end{figure}

Figure~\ref{MagT} plots the transmission at 14 and 131~K as functions of the magnetic field.  
Because the experiment at 131~K was performed with a lower energy for magnetic-field generation, the maximum field was 240~T. 
The transmission at zero field is lower than that at 14~K because the temperature is close to $T_{\textrm{MI}}$. 
The inset shows the change in the transmission ($\Delta  \rm{Trans.}$) as a function of the magnetic field. 
$\Delta  \rm{Trans.}$ is found to become finite at a certain magnetic field ($B^*$), as indicated by the arrow for each temperature.
$B^*$ is evaluated as approximately 120 and 100~T at 14 and 131~K, respectively. 

The optical absorption spectra shown in Fig.~\ref{zero_spectra} and their temperature dependence are similar to those of VO$_2$ \cite{Okazaki2006} 
and in good agreement with previous results for V$_{1-x}$W$_x$O$_2$ thin films \cite{Lee2012}.
The significant temperature dependence in the spectra at energies less than approximately 1.2~eV is due to the change in the energy band structure associated 
with the absorption arising from the  $d_{||} \rightarrow \pi^*$ transition and due to enhancement in free-carrier absorption, called Drude absorption, at high temperatures. 
Because the lowest photon energy in this work is 0.5 eV, the onset of the absorption corresponding to the energy gap is not observed.  
According to previous work \cite{Lee2012}, the energy gap for $x$=0.06 is expected to be approximately 0.1~eV. 

The considerable decrease of optical transmission at 1.977~$\mu$m 
under an ultrahigh magnetic field, as shown in Fig.~\ref{MagT}, constitutes direct evidence of a magnetic-field-induced IM (MFI-IM) transition. 
Because the transmission change is induced at a threshold magnetic field $B^*$, the MFI-IM transition is likely a first-order transition. 
The transition is much broader than expected for a first-order transition because the inhomogeneous distribution of W sites \cite{Tan2012} creates several nanoscale domains with slightly different potential barriers. 

\begin{figure}
\begin{center}
\includegraphics[width=9cm]{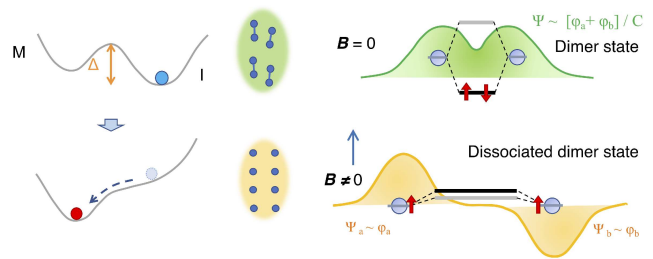}
\caption{\textbf{Schematic of the magnetic-field-induced insulator-metal (MFI-IM) transition.} The left side shows that 
the potential barrier $\Delta$ is lowered because of the Zeeman energy. The middle part schematically shows the collapse of the V--V dimers. 
The right part shows that applying a magnetic field induces the dissociation of the dimer owing to the destabilisation of formation of the bonding state $\Psi$ of 
the molecular orbital, where $\varphi_a$ and $\varphi_b$ are the wave functions of the independent vanadium ions. 
}
\label{chemicata}
\end{center}
\end{figure}

It would not be unreasonable to expect potential barriers between the metallic phase with a uniform V-atom distribution and the insulating phase with 
V--V dimers. 
The dimers are stabilised with the formation of metal--metal bonding, and the dimer can be regarded as a diatomic molecule \cite{Goodenough1960, Hiroi2015}. 
As schematically shown in  Fig.~\ref{chemicata}, one of the most plausible explanations for the observed MFI-IM is that the V--V dimers are dissociated 
by the collapse of the molecular orbital. 
Because the molecular orbital is stable owing to the occupation of two electrons in the bonding sate and their spins need to be antiparallel, making the spins of the two electrons parallel can disturb the formation of bonding between adjacent V atoms. 
The spin quantum number of V$^{4+}$ ($d^1$) is $S$ = 1/2, and its Zeeman energy at 120~T corresponds to 162~K, which is close to the onset temperature for 
the MI transition at zero magnetic field, as shown in Fig.~\ref{rho_trans}. 
This fact supports the interpretation above. 
The potential barrier $\Delta$ in Fig.~\ref{chemicata} for the MFI-IM transition is expected to be as large as the Zeeman energy at 120~T. 
In contrast, the energy separation of the bonding and anti-bonding orbitals of the V--V dimer is expected to be approximately 2.5 eV \cite{Goodenough1971,Eyert2002}. 
Because this energy scale corresponds to 30,000~K, which is more than two orders of magnitude larger than the Zeeman energy, the observed magnetic-field-induced metallisation cannot be explained by considering only an isolated single dimer. 
Some many-body interactions through the electron correlation \cite{Biermann2005,Huffman2017} would be necessary to understand the mechanism of 
dissociation of the dimers by controlling the electron spins. 

Here, it is worth noting how the W doping affects the basic properties of VO$_2$. 
As shown for the V$_{1-x}$W$_x$O$_2$ ($x =0.06$) thin film investigated in this work,
doping the V sublattice with W reduces the MI transition temperature $T_{\textrm{MI}}$ by $\sim$20 K/at.\%W for the bulk \cite{Horlin1972} and 
by $\sim$50 K/at.\%W for nanostructures \cite{Shibuya2010,Wu2011}. 
The microscopic origin of the reduced $T_{\textrm{MI}}$ was revealed by the extended X-ray absorption fine structure as follows. 
The local lattice structure at W sites is more symmetric than that at V sites and induces the detwisting of the nearby asymmetric monoclinic 
VO$_2$ lattice towards the rutile phase \cite{Tan2012}.
The W sites form rutile-like VO$_2$ nuclei, and the propagation of these nuclei decreases the energy barrier of the 
phase transition \cite{Tan2012}. 
Hence, the intrinsic mechanism of the MI transition of W-doped VO$_2$ is identical to that of non-doped VO$_2$. 
The broadening of the transition found in the temperature dependence of the electrical resistivity is considered to be 
due to the inhomogeneous distribution of W sites and reflects the growing so-called ``metallic puddles'' in the insulating host 
\cite{Tan2012,Qazilbash2007,Choi1996}. 
Additionally, the growing part of the metallic puddles gradually reduces the optical transmission. 

Another important issue regarding the reliability of the experimental results is the temperature of the sample in pulsed ultrahigh magnetic fields.  
When the sample is electrically conducting, eddy currents can heat it up. 
Because the pulse duration of the magnetic field in the present work is of the microsecond order, the time derivative of the magnetic field $dB/dt$ 
is large and may induce a significant increase in the sample temperature. 
We evaluated the temperature rise of the sample when it is in the insulating phase and found that the temperature rise is less than 4~K even at 500~T. 
We also found that even after the MFI-IM transition occurs and the resistivity becomes low, the temperature does not increase, because of 
fast thermal relaxation due to the very small thickness of the sample (15~nm). 
The nearly isothermal condition of the sample under a magnetic field has been experimentally proven by using a pulsed magnetic field with different time evolutions (a field  
generated by the single-turn-coil technique \cite{Miura2003}). 
Although the calculated temperature quickly increases immediately after applying the magnetic field and reaches 300~K at 10~T when the sample temperature is 95~K,
the measured optical transmission is almost constant up to approximately 100~T, indicating that no significant eddy-current heating occurs. 
A detailed discussion and evaluation of the sample temperature are presented in the Supplementary Information. 
The sample temperature is maintained even after the MFI-IM transition, and the isothermal process of the magnetic-field-induced phase transition has been 
measured in the present study. 

In summary, we demonstrated that an ultrahigh magnetic field of 500~T can transform the insulating dimerised state of W-doped VO$_2$ to a metallic state at low temperatures. 
We showed that the V--V dimers are dissociated by the collapse of the molecular orbital; the bonding state becomes unstable with the aligning of electron spins 
in the magnetic-field direction. 
The $d$ electrons participating in the metal--metal bonding becomes itinerant after the breaking of the dimers. 
This phenomenon is similar to ``chemical catastrophe'': the chemical bonding is collapsed by a very strong magnetic field 
through the spin Zeeman effect \cite{Date1995,Detmer1997}. 
It had been considered to occur only in cosmic spaces such as on a neutron star, where a very strong magnetic field exceeding 10$^6$~T exists. 
%
The present work suggests that the formation of the molecular orbital between the vanadium ions, which results in the localisation of the unpaired $d$ electrons, is 
the predominant driving force behind the MI transition. 
On the other hand, the Zeeman energy corresponding to 500~T is approximately 60~meV, which is less than the eV-order binding energy for a local isolated V--V dimer by more than two orders of magnitude. 
Therefore, it is likely that the electronic correlation must also be included to obtain a quantitative understanding of the MFI-IM transition \cite{Biermann2005,Huffman2017}.  

%
From the perspective of material science, because 
a singlet spin state with the formation of a cluster of magnetic atoms is exhibited by various other strongly correlated insulators, such as Ti$_2$O$_3$ \cite{Eyert2005,Chang2018}, AlV$_2$O$_4$ \cite{Horibe2006}, 
and CuIr$_2$S$_4$ \cite{Radaelli2002}, 
investigation of the effect of a magnetic field on their electronic states is an intriguing and important research problem. 
A magnetic field of the order of 1000~T is inevitable for such research because of their high $T_{\textrm{MI}}$ of several hundreds of kelvins \cite{Imada1998}.


\section{Methods}
A thin film of V$_{1-x}$W$_x$O$_2$ ($x =0.06$) was prepared using pulsed laser deposition on a TiO$_2$ (001) substrate \cite{Muraoka2002}. 
The film has a 001-oriented single phase, and its thickness is 15~nm.
The optical absorption spectra at zero field and different temperatures are acquired using a commercial spectrometer (JASCO V570) in the transmission configuration. 
The temperature dependence of the electrical resistivity was measured using the conventional four-probe method.  
Magneto-transmission measurements were performed using an infrared fibre laser (AdValue Photonics : AP-TM-1975-SM-05) having 
a wavelength of 1.977 $\mu$m. 
A HgCdTe photodiode was used for detecting the intensity of the transmitted light. 
A 1000~T field generator using electromagnetic flux compression \cite{Nakamura2018} was employed to obtain a strong magnetic field of 520~T.  
The infrared laser and HgCdTe detector were placed in a shield room to avoid electromagnetic noise during the magnetic-field generation. 
The sample has an area of $1.8 \times 1.8 \times$ mm$^2$, and it is sandwiched by two optical fibres having a core diameter of 800 $\mu$m, 
which are used for relaying incoming and transmitted laser light.   
Helium-flow-type cryostats made of plastic were used to achieve low temperatures for the magneto-transmission experiments under ultrahigh magnetic fields. 

\section{Acknowledgements} 
This work was supported by a JSPS KAKENHI Grant-in-Aid for Challenging Exploratory Research under Grant Number 18K18728.

\section{Author contributions} 
Y.H.M. conceptualised the experiment. 
Y.H.M., D.N., A.I., and S.T. performed the optical spectroscopy and ultrahigh-magnetic-field magneto-transmission experiments.
Y.M. and Y.S. performed crystal synthesis and electrical resistivity measurements. 
Y.H.M. performed data analysis and wrote the manuscript with input from all co-authors.

\section{Supplementary Information}


\subsection{Evaluation of heating of the sample in pulsed magnetic fields}
\begin{figure}
\begin{center}
\includegraphics[width=9.5cm]{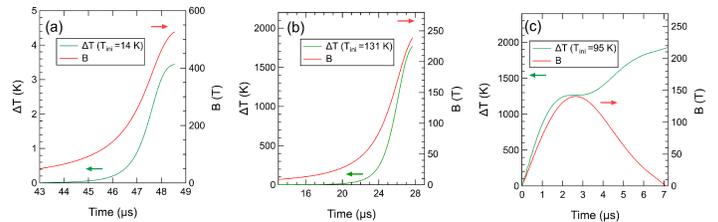}
\caption{Evolution of the temperature rise $\Delta T$ in the V$_{1-x}$W$_x$O$_2$ ($x$ = 0.06) thin film and the $B$ curve 
as a function of time. (a) The initial temperature $T_{\rm{ini}}$ = 14~K and the $B$ is generated by the electromagnetic-flux-compression 
\cite{Nakamura2018}. (b) $T_{\rm{ini}}$ = 131~K and $B$ is generated by the electromagnetic-flux-compression. 
(c) $T_{\rm{ini}}$ = 95~K and $B$ is generated by the single-turn coil technique \cite{Miura2003}. 
}
\label{DeltaT}
\end{center}
\end{figure}

When pulsed magnetic fields are applied to an electrically conducting sample, the sample temperature ($T$) can change with magnetic field ($B$) due to the eddy current heating. 
Here we evaluate the temperature rise ($\Delta T$) of a V$_{1-x}$W$_x$O$_2$ ($x$ = 0.06) thin film in ultrahigh pulsed magnetic fields. 
The $\Delta T$ is calculated as follows, 

\begin{eqnarray}
\Delta T = \frac{x^2}{8} \int_{0}^{t}  \frac{1}{\rho c_V} \left( \frac{dB}{dt} \right)^2 dt
\end{eqnarray}
,where $x$, $c_V$, and $\rho$ are the radius, specific heat, and electrical resistivity of the sample, respectively. 
The calculated $\Delta T$  at different measurement conditions are shown in Figs.~\ref{DeltaT} (a), (b), and (c).
The $x$ is 0.9 mm and $c_V$ is assumed to be $\beta T^3$, where 
$\beta \sim 7.4 \times 10^{-2}$ J/(m$^3$ K$^4$) is obtained from $c_V$ =2~MJ/(m$^3$ K) at 300~K for VO$_2$ \cite{Hamaoui2019}. 
The $\rho$ and $c_V$ used are shown in Table~\ref{para}. %
\begin{table}[hbtp]
  \caption{Parameters use for calculation of the $\Delta T$}
  \label{para}
  \centering
  \begin{tabular}{cccc}
    \hline
     & 14~K  &  95~K  &  131~K  \\
    \hline \hline
   $\rho$ [$\Omega$m]  & 10 & 6 $\times 10^{-5}$ & $3 \times 10^{-6}$ \\
   $c_V$  [J/(m$^3$ K)] & 200   & 6  $\times 10^{4}$ &  $2 \times 10^{5} $ \\
    \hline
  \end{tabular}
\end{table}

As shown in Fig.~\ref{DeltaT} (a), the $\Delta T$ is found to be smaller than 4~K if the initial temperature $T_{\rm{ini}}$ = 14~K even at 500~T.
Hence the significant change in the optical transmission observed in high magnetic fields exceeding 120~T when $T_{\rm{ini}}$ = 14~K 
cannot be attributed to the effect of the eddy current heating. 

On the other hand, the calculated $\Delta T$ for $T_{\rm{ini}}$ = 131~K in $B$ of up to 240~T (Fig.~\ref{DeltaT} (b)) suggests that significant 
heating of the sample takes place. It is because the $\rho$ is rather small reflecting metallic nature. 
The calculated $\Delta T$ reaches 100~K even at a low field of around 60~T.
However, the experimentally obtained optical transmission is found to show nearly no $B$ dependence when the field is 
lower than 100~T, which indicates that the actual $\Delta T$ is smaller than a few Kelvin. 

\begin{figure}
\begin{center}
\includegraphics[width=9cm]{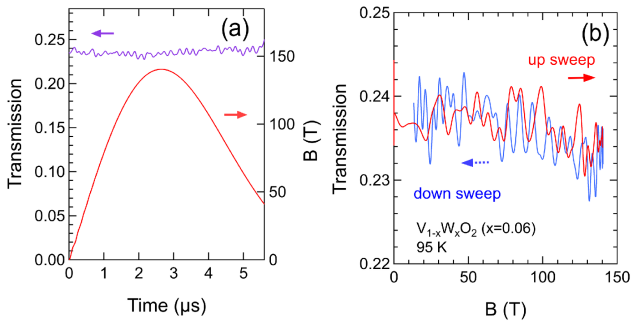}
\caption{(a) Time dependence of $B$ and that of the optical transmission of the V$_{1-x}$W$_x$O$_2$ ($x$ = 0.06) thin film at 1.977~ $\mu$m. 
(b) Plot of the optical transmission as a function of $B$. 
}
\label{STC}
\end{center}
\end{figure}

The similar finding is obtained from another experiment using the single-turn coil technique \cite{Miura2003}. 
The waveform of $B$ is different from the one obtained by electromagnetic flux compression. 
As shown in Fig.~\ref{DeltaT} (c), the sinusoidal like $B$ curve can induce the eddy current heating just after application of field. 
The corresponding optical transmission experiment was conducted on the V$_{1-x}$W$_x$O$_2$ ($x$ = 0.06) thin film at 95~K. 
Fig.~\ref{STC} (a) shows the time evolution of $B$ and that of transmission at 1.977 $\mu$m. 
A significant decrease of the transmission is expected from the calculated $\Delta T$ because the temperature of the sample becomes 300~K at 10~T. 
One find, however, that the transmission keeps the initial value up to around 100~T. 
Moreover, a slight decrease of the transmission observed at field exceeding 100~T (Fig.~\ref{STC} (b)) agrees with the results of higher field 
experiments up to 240 and 520~T. 

From the experimental findings that transmission at 95~ K and 131~K seem to be free from eddy-current heating, we can conclude that 
the calculated $\Delta T$ with adiabatic condition shown in Fig.~\ref{DeltaT} considerably overestimate the effect of sample heating. 
It should be taken into account transferring heat from the sample to the surrounding heat bath. 
In the present study, the heat generated in the sample actually can diffuse to the TiO$_2$ substrate and the heat transfer can be very fast because 
the thickness of the V$_{1-x}$W$_x$O$_2$ ($x$ = 0.06) film ($d$) is only 15~nm. 
Insulating and nonmagnetic TiO$_2$ substrate  of which thickness is 0.5 mm would work as an ideal thermal bath.

The speed of the temperature transportation of the sample with the surface area $A$ and volume $V$ can be evaluated as follows using the Fourier's law,
\begin{eqnarray}
\frac{\partial T}{\partial t} = - \left( \frac{k}{c_V} \right) \frac{A}{V} \frac{\partial T}{\partial x} 
\end{eqnarray}
, where $k$ is the thermal conductivity. 
The speed $\partial T/ \partial t $ is proportional to the gradient of the temperature in space $ \partial T/ \partial x  $. 
Here $A / V = d $= 15 nm, and $k$ can be taken to be 20 W/(m K) for metallic VO$_2$ \cite{Hamaoui2019}. 

To obtain isothermal condition for measurements, it is required to obtain thermal equilibrium condition by  
fast heat exchange with surrounding thermal bath. 
Considering $\Delta T$ = 1~K at the surface of the film, the distance between the surface and the substrate of 15~nm gives the relation 
\begin{eqnarray}
\frac{\partial T}{\partial x} \sim  \frac{\Delta T}{\Delta x} =  \frac{1\;  {\rm{K}}}{15 \times 10^{-9} {\rm{m}}} \sim 6.7 \times 10^7 {\rm{(K/m)}}.
\end{eqnarray}
Then we have, 
\begin{eqnarray}
\frac{\partial T}{\partial t} \sim  - \frac{20}{2 \times 10^6} \left( \frac{1}{15 \times 10^{-9}} \right) (6.7 \times10^7) \nonumber \\
\sim 4.5 \times 10^{10} {\rm{(K/s)}}.
\end{eqnarray}
This is the speed of temperature transfer. Temperature increase of 1~K at the surface can be transferred to the interface between the sample and the TiO$_2$ substrate 
in $(4.5  \times 10^{10} )^{-1}$ s $\sim$ 2.2 $\times 10^{-11}$ s = 22 ps. 
This time scale is six orders of magnitude smaller than the duration time of the magnetic field and thus the isothermal condition is expected to be maintained during the 
$B$ pulse. 
This fast thermal relaxation can explain the experimental findings that the isothermal condition is likely to be maintained during the microsecond ultrahigh $B$ pulse. 
\subsection{Curve fitting of the absorption spectra at different temperatures}
\begin{figure}
\begin{center}
\includegraphics[width=9cm]{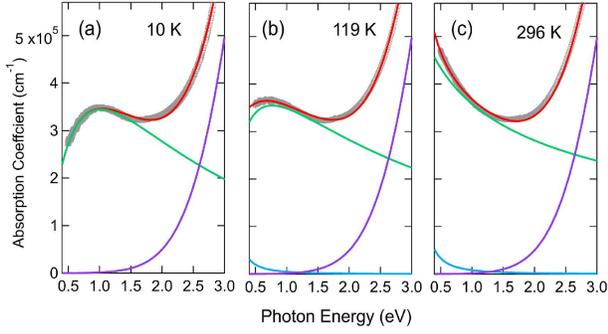}
\caption{Results of the curve fitting (thick solid curves) for the optical absorption spectra of the V$_{1-x}$W$_x$O$_2$ ($x$ = 0.06) thin film 
at different temperatures. The green and purple curves correspond to the $d_{||} \rightarrow \pi^*$ and the CT transitions, respectively. 
The light-blue curve is the Drude component, and the red curve contains the total absorption components.
The experimental results are represented by grey open circles. (a) 10~K. (b) 119~K. (c) 296~K. 
}
\label{spectra}
\end{center}
\end{figure}

The optical absorption spectra of the V$_{1-x}$W$_x$O$_2$ ($x$ = 0.06) thin film are analyzed. In the insulating phase, as shown in Fig.~\ref{spectra} (a), 
the spectrum exhibits a clear absorption band around~1 eV and another absorption rise starts at around 2~eV indicating larger absorption band at higher energy. 
According to the previous studies \cite{Gavini1972,Eyert2002,He2016}, they are the absorption bands due to the $d_{||} \rightarrow \pi^*$ 
and $d_{||} \rightarrow \sigma^*$ transitions, respectively. Here $d_{||}$ is the bonding orbital of the vanadium dimers and $\pi^*$ and $\sigma^*$ are the orbitals originate from $t_{2g}$ and $e_g$  orbitals.  Because the $\sigma^*$ is rather strongly hybridized with oxygen $2p$ orbital, the latter transition can be regarded as a charge transfer (CT) absorption.  
On the other hand, $\pi^*$ has mostly $d_{xz}$ and $d_{yz}$ character of $d$ electrons of a vanadium atom \cite{Eyert2002}. 
A lognormal function is used for representing the  $d_{||} \rightarrow \pi^*$ transition contribution since the peak shape is asymmetric \cite{Okazaki2006, Shibuya2010},
while a Gauss function is used to fit the slope of the the CT transition. 

In addition to the two absorption bands, free carrier absorption (so-called Drude absorption) is taken into account for the spectra fitting. 
The absorption coefficient $\alpha$ for the Drude component is expressed as follows. 
\begin{eqnarray}
\alpha = \frac{\omega_p^2 \lambda^2}{4 \pi c^3 \bar{n} \tau}.
\end{eqnarray}
Here, the plasma frequency $\omega_p$ is proportional to the square root of the carrier density $n_e$. 
\begin{eqnarray}
\omega_p = \sqrt{\frac{n_e e^2}{\varepsilon_0 m^*}}.
\end{eqnarray}
$\lambda$ is the wavelength,  $c$ is the speed of light,  $\bar{n}$ is the reflective index, and $\tau$ is the scattering time. 
$\varepsilon_0$ and $m^*$ are the dielectric constant of vacuum and the effective mass, respectively. 
$m^* = 3 m_0$ \cite{Okazaki2006,Brito2016,Muraoka2018} and $\bar{n}$ =3 \cite{Verleur1968} are used for the fitting, where the $m_0$ is the free electron mass. 

The representative optical absorption spectra at 10, 119, and 296~K are shown in Fig.~\ref{spectra} along with the fitting curves. The red curve is the result of the fitting. 
The green and purple curves are the components of $d_{||} \rightarrow \pi^*$ and the CT transitions, respectively. 
The peak energy of the Lognormal function ($E_0$) seems to change with temperature, suggesting significant change in the electronic structure due to the metal-insulator 
transition of this sample around 100~K. 

\begin{figure}
\begin{center}
\includegraphics[width=9cm]{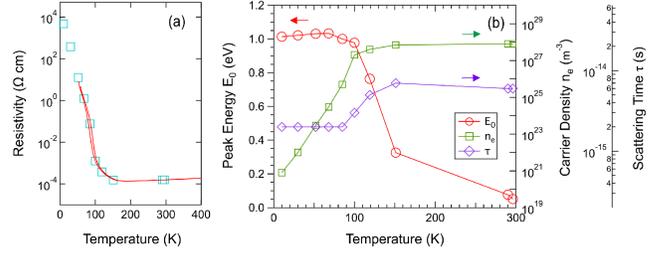}
\caption{(a) The DC electrical resistivity of the V$_{1-x}$W$_x$O$_2$ ($x$ = 0.06) thin film as a function of temperature. 
(b) The deduced fitting parameters as a function of temperature; $E_0$ is the energy peak of the $d_{||} \rightarrow \pi^*$ transition, 
$n_e$ is the carrier density, and the $\tau$ is the scattering time of the carrier. 
}
\label{fitpara}
\end{center}
\end{figure}

%
The Drude term is not significant at temperatures lower than around 70~K at which the $n_e$ is estimated to be around $ 10^{25}$~m$^{-3}$. 
At higher temperatures, the Drude term contributes the optical absorption (light blue curve in Fig.~\ref{spectra}). 
Because there are a lot of adjustable parameters and the slope due to the CT transition seems to be rather independent of temperature,  
we assume that the CT transition does not depend on temperature. 
Regarding the Drude absorption, we tried to find $n_e$ and $\tau$ that give a good fitting results and simultaneously explains the DC electrical 
resistivity shown in Fig.~\ref{fitpara} (a). 
%
%
The DC electrical resistivity $\rho$ is assumed to be expressed as follows.
\begin{eqnarray}
\rho = \frac {m^*}{e^2 n_e \tau}.
\end{eqnarray}
The light-blue open squares shown in Fig.~\ref{fitpara} (a) are the evaluated $\rho$ with the parameters used for the curve fitting for the 
optical absorption spectra. 
The parameters used are shown in Fig.~\ref{fitpara} (b) as a function of temperature. 
Although the Drude theorem can be too simple to evaluate the electronic state of V$_{1-x}$W$_x$O$_2$ ($x$ = 0.06), the obtained $n_e$ at 
higher temperature around 10$^{28}$ m$^{-3}$ is in agreement with the expected carrier density 1.7 $\times$ 10$^{28}$ m$^{-3}$ assuming one 
electron per formula unit of VO$_2$. (Here we use the density 2.33 g/cm$^3$ and effects of W-doping is not taken into account.) 
Because the scattering time of carriers in metallic VO$_2$ can be estimated as an order of $10^{-15}  \sim  10^{-14}$  s$^{-1}$ \cite{Okazaki2006}, 
the $\tau$ shown in Fig.~\ref{fitpara} (b) also seem be rather reasonable values . 

As shown in the results of the curve fitting (Fig.~\ref{spectra}), it is found that the contribution to the absorption at 0.627~eV (1.977 $\mu$ m) in the spectra of V$_{1-x}$W$_x$O$_2$ ($x$ = 0.06)  mainly comes 
from the absorption due to the $d_{||} \rightarrow \pi^*$ transition with a small contribution of the free carrier absorption. 
Therefore, the observed significant decrease of the transmission at 1.977 $\mu$m in the ultrahigh magnetic fields exceeding 100~T to 520~T can be attributed to the 
change in the electronic state. 
The $d_{||} \rightarrow \pi^*$ absorption band shifts to the lower energy with magnetic field and most probably close the energy gap at around 500~T.

\bibliography{wvo}

\begin{thebibliography}{36}%
\makeatletter
\providecommand \@ifxundefined [1]{%
 \@ifx{#1\undefined}
}%
\providecommand \@ifnum [1]{%
 \ifnum #1\expandafter \@firstoftwo
 \else \expandafter \@secondoftwo
 \fi
}%
\providecommand \@ifx [1]{%
 \ifx #1\expandafter \@firstoftwo
 \else \expandafter \@secondoftwo
 \fi
}%
\providecommand \natexlab [1]{#1}%
\providecommand \enquote  [1]{``#1''}%
\providecommand \bibnamefont  [1]{#1}%
\providecommand \bibfnamefont [1]{#1}%
\providecommand \citenamefont [1]{#1}%
\providecommand \href@noop [0]{\@secondoftwo}%
\providecommand \href [0]{\begingroup \@sanitize@url \@href}%
\providecommand \@href[1]{\@@startlink{#1}\@@href}%
\providecommand \@@href[1]{\endgroup#1\@@endlink}%
\providecommand \@sanitize@url [0]{\catcode `\\12\catcode `\$12\catcode
  `\&12\catcode `\#12\catcode `\^12\catcode `\_12\catcode `\%12\relax}%
\providecommand \@@startlink[1]{}%
\providecommand \@@endlink[0]{}%
\providecommand \url  [0]{\begingroup\@sanitize@url \@url }%
\providecommand \@url [1]{\endgroup\@href {#1}{\urlprefix }}%
\providecommand \urlprefix  [0]{URL }%
\providecommand \Eprint [0]{\href }%
\providecommand \doibase [0]{https://doi.org/}%
\providecommand \selectlanguage [0]{\@gobble}%
\providecommand \bibinfo  [0]{\@secondoftwo}%
\providecommand \bibfield  [0]{\@secondoftwo}%
\providecommand \translation [1]{[#1]}%
\providecommand \BibitemOpen [0]{}%
\providecommand \bibitemStop [0]{}%
\providecommand \bibitemNoStop [0]{.\EOS\space}%
\providecommand \EOS [0]{\spacefactor3000\relax}%
\providecommand \BibitemShut  [1]{\csname bibitem#1\endcsname}%
\let\auto@bib@innerbib\@empty
\bibitem [{\citenamefont {Morin}(1959)}]{Morin1959}%
  \BibitemOpen
  \bibfield  {author} {\bibinfo {author} {\bibfnamefont {F.~J.}\ \bibnamefont
  {Morin}},\ }\bibfield  {title} {\bibinfo {title} {Oxides which show a
  metal-to-insulator transition at the neel temperature},\ }\href
  {https://doi.org/10.1103/PhysRevLett.3.34} {\bibfield  {journal} {\bibinfo
  {journal} {Phys. Rev. Lett.}\ }\textbf {\bibinfo {volume} {3}},\ \bibinfo
  {pages} {34} (\bibinfo {year} {1959})}\BibitemShut {NoStop}%
\bibitem [{\citenamefont {Zylbersztejn}\ and\ \citenamefont
  {Mott}(1975)}]{Zylbersztejn1975}%
  \BibitemOpen
  \bibfield  {author} {\bibinfo {author} {\bibfnamefont {A.}~\bibnamefont
  {Zylbersztejn}}\ and\ \bibinfo {author} {\bibfnamefont {N.~F.}\ \bibnamefont
  {Mott}},\ }\bibfield  {title} {\bibinfo {title} {Metal-insulator transition
  in vanadium dioxide},\ }\href {https://doi.org/10.1103/PhysRevB.11.4383}
  {\bibfield  {journal} {\bibinfo  {journal} {Phys. Rev. B}\ }\textbf {\bibinfo
  {volume} {11}},\ \bibinfo {pages} {4383} (\bibinfo {year}
  {1975})}\BibitemShut {NoStop}%
\bibitem [{\citenamefont {Rice}\ \emph {et~al.}(1994)\citenamefont {Rice},
  \citenamefont {Launois},\ and\ \citenamefont {Pouget}}]{Rice1994}%
  \BibitemOpen
  \bibfield  {author} {\bibinfo {author} {\bibfnamefont {T.~M.}\ \bibnamefont
  {Rice}}, \bibinfo {author} {\bibfnamefont {H.}~\bibnamefont {Launois}},\ and\
  \bibinfo {author} {\bibfnamefont {J.~P.}\ \bibnamefont {Pouget}},\ }\bibfield
   {title} {\bibinfo {title} {Comment on "vo${_2}$: Peierls or mott-hubbard? a
  view from band theory"},\ }\href
  {https://doi.org/10.1103/PhysRevLett.73.3042} {\bibfield  {journal} {\bibinfo
   {journal} {Phys. Rev. Lett.}\ }\textbf {\bibinfo {volume} {73}},\ \bibinfo
  {pages} {3042} (\bibinfo {year} {1994})}\BibitemShut {NoStop}%
\bibitem [{\citenamefont {Huffman}\ \emph {et~al.}(2017)\citenamefont
  {Huffman}, \citenamefont {Hendriks}, \citenamefont {Walter}, \citenamefont
  {Yoon}, \citenamefont {Ju}, \citenamefont {Smith}, \citenamefont {Carr},
  \citenamefont {Krakauer},\ and\ \citenamefont {Qazilbash}}]{Huffman2017}%
  \BibitemOpen
  \bibfield  {author} {\bibinfo {author} {\bibfnamefont {T.~J.}\ \bibnamefont
  {Huffman}}, \bibinfo {author} {\bibfnamefont {C.}~\bibnamefont {Hendriks}},
  \bibinfo {author} {\bibfnamefont {E.~J.}\ \bibnamefont {Walter}}, \bibinfo
  {author} {\bibfnamefont {J.}~\bibnamefont {Yoon}}, \bibinfo {author}
  {\bibfnamefont {H.}~\bibnamefont {Ju}}, \bibinfo {author} {\bibfnamefont
  {R.}~\bibnamefont {Smith}}, \bibinfo {author} {\bibfnamefont {G.~L.}\
  \bibnamefont {Carr}}, \bibinfo {author} {\bibfnamefont {H.}~\bibnamefont
  {Krakauer}},\ and\ \bibinfo {author} {\bibfnamefont {M.~M.}\ \bibnamefont
  {Qazilbash}},\ }\bibfield  {title} {\bibinfo {title} {Insulating phases of
  vanadium dioxide are mott-hubbard insulators},\ }\href
  {https://doi.org/10.1103/PhysRevB.95.075125} {\bibfield  {journal} {\bibinfo
  {journal} {Phys. Rev. B}\ }\textbf {\bibinfo {volume} {95}},\ \bibinfo
  {pages} {075125} (\bibinfo {year} {2017})}\BibitemShut {NoStop}%
\bibitem [{\citenamefont {Goodenough}(1960)}]{Goodenough1960}%
  \BibitemOpen
  \bibfield  {author} {\bibinfo {author} {\bibfnamefont {J.~B.}\ \bibnamefont
  {Goodenough}},\ }\bibfield  {title} {\bibinfo {title} {Direct cation- -cation
  interactions in several oxides},\ }\href
  {https://doi.org/10.1103/PhysRev.117.1442} {\bibfield  {journal} {\bibinfo
  {journal} {Phys. Rev.}\ }\textbf {\bibinfo {volume} {117}},\ \bibinfo {pages}
  {1442} (\bibinfo {year} {1960})}\BibitemShut {NoStop}%
\bibitem [{\citenamefont {Wentzcovitch}\ \emph {et~al.}(1994)\citenamefont
  {Wentzcovitch}, \citenamefont {Schulz},\ and\ \citenamefont
  {Allen}}]{Wentzcovitch1994}%
  \BibitemOpen
  \bibfield  {author} {\bibinfo {author} {\bibfnamefont {R.~M.}\ \bibnamefont
  {Wentzcovitch}}, \bibinfo {author} {\bibfnamefont {W.~W.}\ \bibnamefont
  {Schulz}},\ and\ \bibinfo {author} {\bibfnamefont {P.~B.}\ \bibnamefont
  {Allen}},\ }\bibfield  {title} {\bibinfo {title} {${\mathrm{vo}}_{2}$:
  Peierls or mott-hubbard? a view from band theory},\ }\href
  {https://doi.org/10.1103/PhysRevLett.72.3389} {\bibfield  {journal} {\bibinfo
   {journal} {Phys. Rev. Lett.}\ }\textbf {\bibinfo {volume} {72}},\ \bibinfo
  {pages} {3389} (\bibinfo {year} {1994})}\BibitemShut {NoStop}%
\bibitem [{\citenamefont {Eyert}(2002)}]{Eyert2002}%
  \BibitemOpen
  \bibfield  {author} {\bibinfo {author} {\bibfnamefont {V.}~\bibnamefont
  {Eyert}},\ }\bibfield  {title} {\bibinfo {title} {The metal-insulator
  transitions of v${\mathrm{o}}_{2}$: A band theoretical approach},\ }\href
  {https://doi.org/10.1002/1521-3889(200210)11:9<650::AID-ANDP650>3.0.CO;2-K}
  {\bibfield  {journal} {\bibinfo  {journal} {Annalen der Physik}\ }\textbf
  {\bibinfo {volume} {11}},\ \bibinfo {pages} {650} (\bibinfo {year}
  {2002})}\BibitemShut {NoStop}%
\bibitem [{\citenamefont {Hiroi}(2015)}]{Hiroi2015}%
  \BibitemOpen
  \bibfield  {author} {\bibinfo {author} {\bibfnamefont {A.}~\bibnamefont
  {Hiroi}},\ }\bibfield  {title} {\bibinfo {title} {Structural instability of
  the rutile compounds and its relevance to the metal-insulator transition of
  vo2},\ }\href
  {https://doi.org/https://doi.org/10.1016/j.progsolidstchem.2015.02.001}
  {\bibfield  {journal} {\bibinfo  {journal} {Progress in Solid State
  Chemistry}\ }\textbf {\bibinfo {volume} {43}},\ \bibinfo {pages} {47}
  (\bibinfo {year} {2015})}\BibitemShut {NoStop}%
\bibitem [{\citenamefont {Imada}\ \emph {et~al.}(1998)\citenamefont {Imada},
  \citenamefont {Fujimori},\ and\ \citenamefont {Tokura}}]{Imada1998}%
  \BibitemOpen
  \bibfield  {author} {\bibinfo {author} {\bibfnamefont {M.}~\bibnamefont
  {Imada}}, \bibinfo {author} {\bibfnamefont {A.}~\bibnamefont {Fujimori}},\
  and\ \bibinfo {author} {\bibfnamefont {Y.}~\bibnamefont {Tokura}},\
  }\bibfield  {title} {\bibinfo {title} {Metal-insulator transitions},\ }\href
  {https://doi.org/10.1103/RevModPhys.70.1039} {\bibfield  {journal} {\bibinfo
  {journal} {Rev. Mod. Phys.}\ }\textbf {\bibinfo {volume} {70}},\ \bibinfo
  {pages} {1039} (\bibinfo {year} {1998})}\BibitemShut {NoStop}%
\bibitem [{\citenamefont {Biermann}\ \emph {et~al.}(2005)\citenamefont
  {Biermann}, \citenamefont {Poteryaev}, \citenamefont {Lichtenstein},\ and\
  \citenamefont {Georges}}]{Biermann2005}%
  \BibitemOpen
  \bibfield  {author} {\bibinfo {author} {\bibfnamefont {S.}~\bibnamefont
  {Biermann}}, \bibinfo {author} {\bibfnamefont {A.}~\bibnamefont {Poteryaev}},
  \bibinfo {author} {\bibfnamefont {A.~I.}\ \bibnamefont {Lichtenstein}},\ and\
  \bibinfo {author} {\bibfnamefont {A.}~\bibnamefont {Georges}},\ }\bibfield
  {title} {\bibinfo {title} {Dynamical singlets and correlation-assisted
  peierls transition in ${\mathrm{v}\mathrm{o}}_{2}$},\ }\href
  {https://doi.org/10.1103/PhysRevLett.94.026404} {\bibfield  {journal}
  {\bibinfo  {journal} {Phys. Rev. Lett.}\ }\textbf {\bibinfo {volume} {94}},\
  \bibinfo {pages} {026404} (\bibinfo {year} {2005})}\BibitemShut {NoStop}%
\bibitem [{\citenamefont {Najera}\ \emph {et~al.}(2017)\citenamefont {Najera},
  \citenamefont {Civelli}, \citenamefont {Dobrosavljevic},\ and\ \citenamefont
  {Rozenberg}}]{Najera2017}%
  \BibitemOpen
  \bibfield  {author} {\bibinfo {author} {\bibfnamefont {O.}~\bibnamefont
  {Najera}}, \bibinfo {author} {\bibfnamefont {M.}~\bibnamefont {Civelli}},
  \bibinfo {author} {\bibfnamefont {V.}~\bibnamefont {Dobrosavljevic}},\ and\
  \bibinfo {author} {\bibfnamefont {M.~J.}\ \bibnamefont {Rozenberg}},\
  }\bibfield  {title} {\bibinfo {title} {Resolving the ${\mathrm{vo}}_{2}$
  controversy: Mott mechanism dominates the insulator-to-metal transition},\
  }\href {https://doi.org/10.1103/PhysRevB.95.035113} {\bibfield  {journal}
  {\bibinfo  {journal} {Phys. Rev. B}\ }\textbf {\bibinfo {volume} {95}},\
  \bibinfo {pages} {035113} (\bibinfo {year} {2017})}\BibitemShut {NoStop}%
\bibitem [{\citenamefont {Jia}\ \emph {et~al.}(2018)\citenamefont {Jia},
  \citenamefont {Shu}, \citenamefont {Gao}, \citenamefont {Cheng},
  \citenamefont {Peng}, \citenamefont {Fan}, \citenamefont {Liu},\ and\
  \citenamefont {Wang}}]{Jia2018}%
  \BibitemOpen
  \bibfield  {author} {\bibinfo {author} {\bibfnamefont {Z.-Y.}\ \bibnamefont
  {Jia}}, \bibinfo {author} {\bibfnamefont {F.-Z.}\ \bibnamefont {Shu}},
  \bibinfo {author} {\bibfnamefont {Y.-J.}\ \bibnamefont {Gao}}, \bibinfo
  {author} {\bibfnamefont {F.}~\bibnamefont {Cheng}}, \bibinfo {author}
  {\bibfnamefont {R.-W.}\ \bibnamefont {Peng}}, \bibinfo {author}
  {\bibfnamefont {R.-H.}\ \bibnamefont {Fan}}, \bibinfo {author} {\bibfnamefont
  {Y.}~\bibnamefont {Liu}},\ and\ \bibinfo {author} {\bibfnamefont
  {M.}~\bibnamefont {Wang}},\ }\bibfield  {title} {\bibinfo {title}
  {Dynamically switching the polarization state of light based on the phase
  transition of vanadium dioxide},\ }\href
  {https://doi.org/10.1103/PhysRevApplied.9.034009} {\bibfield  {journal}
  {\bibinfo  {journal} {Phys. Rev. Applied}\ }\textbf {\bibinfo {volume} {9}},\
  \bibinfo {pages} {034009} (\bibinfo {year} {2018})}\BibitemShut {NoStop}%
\bibitem [{\citenamefont {Shibuya}\ \emph {et~al.}(2010)\citenamefont
  {Shibuya}, \citenamefont {Kawasaki},\ and\ \citenamefont
  {Tokura}}]{Shibuya2010}%
  \BibitemOpen
  \bibfield  {author} {\bibinfo {author} {\bibfnamefont {K.}~\bibnamefont
  {Shibuya}}, \bibinfo {author} {\bibfnamefont {M.}~\bibnamefont {Kawasaki}},\
  and\ \bibinfo {author} {\bibfnamefont {Y.}~\bibnamefont {Tokura}},\
  }\bibfield  {title} {\bibinfo {title} {Metal-insulator transition in
  epitaxial ${\mathrm{v}}_{1-x} {\mathrm{w}}_{x} {\mathrm{o}}_{2}$ (${0 \leq x
  \leq 0.33}$) thin films},\ }\href {https://doi.org/10.1063/1.3291053}
  {\bibfield  {journal} {\bibinfo  {journal} {Applied Physics Letters}\
  }\textbf {\bibinfo {volume} {96}},\ \bibinfo {pages} {022102} (\bibinfo
  {year} {2010})}\BibitemShut {NoStop}%
\bibitem [{\citenamefont {Gavini}\ and\ \citenamefont
  {Kwan}(1972)}]{Gavini1972}%
  \BibitemOpen
  \bibfield  {author} {\bibinfo {author} {\bibfnamefont {A.}~\bibnamefont
  {Gavini}}\ and\ \bibinfo {author} {\bibfnamefont {C.~C.~Y.}\ \bibnamefont
  {Kwan}},\ }\bibfield  {title} {\bibinfo {title} {Optical properties of
  semiconducting ${\mathrm{v}\mathrm{o}}_{2}$ films},\ }\href
  {https://doi.org/10.1103/PhysRevB.5.3138} {\bibfield  {journal} {\bibinfo
  {journal} {Phys. Rev. B}\ }\textbf {\bibinfo {volume} {5}},\ \bibinfo {pages}
  {3138} (\bibinfo {year} {1972})}\BibitemShut {NoStop}%
\bibitem [{\citenamefont {Okazaki}\ \emph {et~al.}(2006)\citenamefont
  {Okazaki}, \citenamefont {Sugai}, \citenamefont {Muraoka},\ and\
  \citenamefont {Hiroi}}]{Okazaki2006}%
  \BibitemOpen
  \bibfield  {author} {\bibinfo {author} {\bibfnamefont {K.}~\bibnamefont
  {Okazaki}}, \bibinfo {author} {\bibfnamefont {S.}~\bibnamefont {Sugai}},
  \bibinfo {author} {\bibfnamefont {Y.}~\bibnamefont {Muraoka}},\ and\ \bibinfo
  {author} {\bibfnamefont {Z.}~\bibnamefont {Hiroi}},\ }\bibfield  {title}
  {\bibinfo {title} {Role of electron-electron and electron-phonon interaction
  effects in the optical conductivity of ${\mathrm{vo}}_{2}$},\ }\href
  {https://doi.org/10.1103/PhysRevB.73.165116} {\bibfield  {journal} {\bibinfo
  {journal} {Phys. Rev. B}\ }\textbf {\bibinfo {volume} {73}},\ \bibinfo
  {pages} {165116} (\bibinfo {year} {2006})}\BibitemShut {NoStop}%
\bibitem [{\citenamefont {He}\ \emph {et~al.}(2016)\citenamefont {He},
  \citenamefont {Gray}, \citenamefont {Granitzka}, \citenamefont {Jeong},
  \citenamefont {Aetukuri}, \citenamefont {Kukreja}, \citenamefont {Miao},
  \citenamefont {Breitweiser}, \citenamefont {Wu}, \citenamefont {Huang},
  \citenamefont {Olalde-Velasco}, \citenamefont {Pelliciari}, \citenamefont
  {Schlotter}, \citenamefont {Arenholz}, \citenamefont {Schmitt}, \citenamefont
  {Samant}, \citenamefont {Parkin}, \citenamefont {D\"urr},\ and\ \citenamefont
  {Wray}}]{He2016}%
  \BibitemOpen
  \bibfield  {author} {\bibinfo {author} {\bibfnamefont {H.}~\bibnamefont
  {He}}, \bibinfo {author} {\bibfnamefont {A.~X.}\ \bibnamefont {Gray}},
  \bibinfo {author} {\bibfnamefont {P.}~\bibnamefont {Granitzka}}, \bibinfo
  {author} {\bibfnamefont {J.~W.}\ \bibnamefont {Jeong}}, \bibinfo {author}
  {\bibfnamefont {N.~P.}\ \bibnamefont {Aetukuri}}, \bibinfo {author}
  {\bibfnamefont {R.}~\bibnamefont {Kukreja}}, \bibinfo {author} {\bibfnamefont
  {L.}~\bibnamefont {Miao}}, \bibinfo {author} {\bibfnamefont {S.~A.}\
  \bibnamefont {Breitweiser}}, \bibinfo {author} {\bibfnamefont
  {J.}~\bibnamefont {Wu}}, \bibinfo {author} {\bibfnamefont {Y.~B.}\
  \bibnamefont {Huang}}, \bibinfo {author} {\bibfnamefont {P.}~\bibnamefont
  {Olalde-Velasco}}, \bibinfo {author} {\bibfnamefont {J.}~\bibnamefont
  {Pelliciari}}, \bibinfo {author} {\bibfnamefont {W.~F.}\ \bibnamefont
  {Schlotter}}, \bibinfo {author} {\bibfnamefont {E.}~\bibnamefont {Arenholz}},
  \bibinfo {author} {\bibfnamefont {T.}~\bibnamefont {Schmitt}}, \bibinfo
  {author} {\bibfnamefont {M.~G.}\ \bibnamefont {Samant}}, \bibinfo {author}
  {\bibfnamefont {S.~S.~P.}\ \bibnamefont {Parkin}}, \bibinfo {author}
  {\bibfnamefont {H.~A.}\ \bibnamefont {D\"urr}},\ and\ \bibinfo {author}
  {\bibfnamefont {L.~A.}\ \bibnamefont {Wray}},\ }\bibfield  {title} {\bibinfo
  {title} {Measurement of collective excitations in
  ${\mathrm{v}\mathrm{o}}_{2}$ by resonant inelastic x-ray scattering},\ }\href
  {https://doi.org/10.1103/PhysRevB.94.161119} {\bibfield  {journal} {\bibinfo
  {journal} {Phys. Rev. B}\ }\textbf {\bibinfo {volume} {94}},\ \bibinfo
  {pages} {161119} (\bibinfo {year} {2016})}\BibitemShut {NoStop}%
\bibitem [{\citenamefont {Lee}\ \emph {et~al.}(2012)\citenamefont {Lee},
  \citenamefont {Shibuya}, \citenamefont {Kawasaki},\ and\ \citenamefont
  {Tokura}}]{Lee2012}%
  \BibitemOpen
  \bibfield  {author} {\bibinfo {author} {\bibfnamefont {J.~S.}\ \bibnamefont
  {Lee}}, \bibinfo {author} {\bibfnamefont {K.}~\bibnamefont {Shibuya}},
  \bibinfo {author} {\bibfnamefont {M.}~\bibnamefont {Kawasaki}},\ and\
  \bibinfo {author} {\bibfnamefont {Y.}~\bibnamefont {Tokura}},\ }\bibfield
  {title} {\bibinfo {title} {Optical investigation of metal-insulator
  transitions in v${}_{1\ensuremath{-}x}$w${}_{x}$o${}_{2}$ (0
  $\ensuremath{\le}$ $x$ $\ensuremath{\le}$ 0.33)},\ }\href
  {https://doi.org/10.1103/PhysRevB.85.155110} {\bibfield  {journal} {\bibinfo
  {journal} {Phys. Rev. B}\ }\textbf {\bibinfo {volume} {85}},\ \bibinfo
  {pages} {155110} (\bibinfo {year} {2012})}\BibitemShut {NoStop}%
\bibitem [{\citenamefont {Nakamura}\ \emph {et~al.}(2018)\citenamefont
  {Nakamura}, \citenamefont {Ikeda}, \citenamefont {Sawabe}, \citenamefont
  {Matsuda},\ and\ \citenamefont {Takeyama}}]{Nakamura2018}%
  \BibitemOpen
  \bibfield  {author} {\bibinfo {author} {\bibfnamefont {D.}~\bibnamefont
  {Nakamura}}, \bibinfo {author} {\bibfnamefont {A.}~\bibnamefont {Ikeda}},
  \bibinfo {author} {\bibfnamefont {H.}~\bibnamefont {Sawabe}}, \bibinfo
  {author} {\bibfnamefont {Y.~H.}\ \bibnamefont {Matsuda}},\ and\ \bibinfo
  {author} {\bibfnamefont {S.}~\bibnamefont {Takeyama}},\ }\bibfield  {title}
  {\bibinfo {title} {Record indoor magnetic field of 1200 t generated by
  electromagnetic flux-compression},\ }\href
  {https://doi.org/10.1063/1.5044557} {\bibfield  {journal} {\bibinfo
  {journal} {Review of Scientific Instruments}\ }\textbf {\bibinfo {volume}
  {89}},\ \bibinfo {pages} {095106} (\bibinfo {year} {2018})}\BibitemShut
  {NoStop}%
\bibitem [{\citenamefont {Tan}\ \emph {et~al.}(2012)\citenamefont {Tan},
  \citenamefont {Yao}, \citenamefont {Long}, \citenamefont {Sun}, \citenamefont
  {Feng}, \citenamefont {Cheng}, \citenamefont {Yuan}, \citenamefont {Zhang},
  \citenamefont {Liu}, \citenamefont {Wu},\ and\ \citenamefont {andShiqiang
  Wei}}]{Tan2012}%
  \BibitemOpen
  \bibfield  {author} {\bibinfo {author} {\bibfnamefont {X.}~\bibnamefont
  {Tan}}, \bibinfo {author} {\bibfnamefont {T.}~\bibnamefont {Yao}}, \bibinfo
  {author} {\bibfnamefont {R.}~\bibnamefont {Long}}, \bibinfo {author}
  {\bibfnamefont {Z.}~\bibnamefont {Sun}}, \bibinfo {author} {\bibfnamefont
  {Y.}~\bibnamefont {Feng}}, \bibinfo {author} {\bibfnamefont {H.}~\bibnamefont
  {Cheng}}, \bibinfo {author} {\bibfnamefont {X.}~\bibnamefont {Yuan}},
  \bibinfo {author} {\bibfnamefont {W.}~\bibnamefont {Zhang}}, \bibinfo
  {author} {\bibfnamefont {Q.}~\bibnamefont {Liu}}, \bibinfo {author}
  {\bibfnamefont {C.}~\bibnamefont {Wu}},\ and\ \bibinfo {author}
  {\bibfnamefont {Y.~X.}\ \bibnamefont {andShiqiang Wei}},\ }\bibfield  {title}
  {\bibinfo {title} {Unraveling metal-insulator transition mechanism of
  ${\mathrm{v}\mathrm{o}}_{2}$ triggered by tungsten doping},\ }\href@noop {}
  {\bibfield  {journal} {\bibinfo  {journal} {Scientific Reports}\ }\textbf
  {\bibinfo {volume} {2}},\ \bibinfo {pages} {466} (\bibinfo {year}
  {2012})}\BibitemShut {NoStop}%
\bibitem [{\citenamefont {Goodenough}(1971)}]{Goodenough1971}%
  \BibitemOpen
  \bibfield  {author} {\bibinfo {author} {\bibfnamefont {J.~B.}\ \bibnamefont
  {Goodenough}},\ }\bibfield  {title} {\bibinfo {title} {The two components of
  the crystallographic transition in vo2},\ }\href
  {https://doi.org/https://doi.org/10.1016/0022-4596(71)90091-0} {\bibfield
  {journal} {\bibinfo  {journal} {Journal of Solid State Chemistry}\ }\textbf
  {\bibinfo {volume} {3}},\ \bibinfo {pages} {490 } (\bibinfo {year}
  {1971})}\BibitemShut {NoStop}%
\bibitem [{\citenamefont {Horlin}\ \emph {et~al.}(1972)\citenamefont {Horlin},
  \citenamefont {Niklewski},\ and\ \citenamefont {Nygren}}]{Horlin1972}%
  \BibitemOpen
  \bibfield  {author} {\bibinfo {author} {\bibfnamefont {T.}~\bibnamefont
  {Horlin}}, \bibinfo {author} {\bibfnamefont {T.}~\bibnamefont {Niklewski}},\
  and\ \bibinfo {author} {\bibfnamefont {M.}~\bibnamefont {Nygren}},\
  }\bibfield  {title} {\bibinfo {title} {Electrical and magnetic properties of
  v${}_{1\ensuremath{-}x}$w${}_{x}$o${}_{2}$, $0 \leq x \leq0.060$},\ }\href
  {https://doi.org/https://doi.org/10.1016/0025-5408(72)90189-4} {\bibfield
  {journal} {\bibinfo  {journal} {Materials Research Bulletin}\ }\textbf
  {\bibinfo {volume} {7}},\ \bibinfo {pages} {1515 } (\bibinfo {year}
  {1972})}\BibitemShut {NoStop}%
\bibitem [{\citenamefont {Wu}\ \emph {et~al.}(2011)\citenamefont {Wu},
  \citenamefont {Whittaker}, \citenamefont {Banerjee},\ and\ \citenamefont
  {Sambandamurthy}}]{Wu2011}%
  \BibitemOpen
  \bibfield  {author} {\bibinfo {author} {\bibfnamefont {T.-L.}\ \bibnamefont
  {Wu}}, \bibinfo {author} {\bibfnamefont {L.}~\bibnamefont {Whittaker}},
  \bibinfo {author} {\bibfnamefont {S.}~\bibnamefont {Banerjee}},\ and\
  \bibinfo {author} {\bibfnamefont {G.}~\bibnamefont {Sambandamurthy}},\
  }\bibfield  {title} {\bibinfo {title} {Temperature and voltage driven tunable
  metal-insulator transition in individual
  ${\mathrm{w}}_{x}$v${}_{1\ensuremath{-}x}$o${}_{2}$ nanowires},\ }\href
  {https://doi.org/10.1103/PhysRevB.83.073101} {\bibfield  {journal} {\bibinfo
  {journal} {Phys. Rev. B}\ }\textbf {\bibinfo {volume} {83}},\ \bibinfo
  {pages} {073101} (\bibinfo {year} {2011})}\BibitemShut {NoStop}%
\bibitem [{\citenamefont {Qazilbash}\ \emph {et~al.}(2007)\citenamefont
  {Qazilbash}, \citenamefont {Brehm}, \citenamefont {Chae}, \citenamefont {Ho},
  \citenamefont {Andreev}, \citenamefont {Kim}, \citenamefont {Yun},
  \citenamefont {Balatsky}, \citenamefont {Maple}, \citenamefont {Keilmann},
  \citenamefont {Kim},\ and\ \citenamefont {Basov}}]{Qazilbash2007}%
  \BibitemOpen
  \bibfield  {author} {\bibinfo {author} {\bibfnamefont {M.~M.}\ \bibnamefont
  {Qazilbash}}, \bibinfo {author} {\bibfnamefont {M.}~\bibnamefont {Brehm}},
  \bibinfo {author} {\bibfnamefont {B.-G.}\ \bibnamefont {Chae}}, \bibinfo
  {author} {\bibfnamefont {P.-C.}\ \bibnamefont {Ho}}, \bibinfo {author}
  {\bibfnamefont {G.~O.}\ \bibnamefont {Andreev}}, \bibinfo {author}
  {\bibfnamefont {B.-J.}\ \bibnamefont {Kim}}, \bibinfo {author} {\bibfnamefont
  {S.~J.}\ \bibnamefont {Yun}}, \bibinfo {author} {\bibfnamefont {A.~V.}\
  \bibnamefont {Balatsky}}, \bibinfo {author} {\bibfnamefont {M.~B.}\
  \bibnamefont {Maple}}, \bibinfo {author} {\bibfnamefont {F.}~\bibnamefont
  {Keilmann}}, \bibinfo {author} {\bibfnamefont {H.-T.}\ \bibnamefont {Kim}},\
  and\ \bibinfo {author} {\bibfnamefont {D.~N.}\ \bibnamefont {Basov}},\
  }\bibfield  {title} {\bibinfo {title} {Mott transition in vo2 revealed by
  infrared spectroscopy and nano-imaging},\ }\href
  {https://doi.org/10.1126/science.1150124} {\bibfield  {journal} {\bibinfo
  {journal} {Science}\ }\textbf {\bibinfo {volume} {318}},\ \bibinfo {pages}
  {1750} (\bibinfo {year} {2007})}\BibitemShut {NoStop}%
\bibitem [{\citenamefont {Choi}\ \emph {et~al.}(1996)\citenamefont {Choi},
  \citenamefont {Ahn}, \citenamefont {Jung}, \citenamefont {Noh},\ and\
  \citenamefont {Kim}}]{Choi1996}%
  \BibitemOpen
  \bibfield  {author} {\bibinfo {author} {\bibfnamefont {H.~S.}\ \bibnamefont
  {Choi}}, \bibinfo {author} {\bibfnamefont {J.~S.}\ \bibnamefont {Ahn}},
  \bibinfo {author} {\bibfnamefont {J.~H.}\ \bibnamefont {Jung}}, \bibinfo
  {author} {\bibfnamefont {T.~W.}\ \bibnamefont {Noh}},\ and\ \bibinfo {author}
  {\bibfnamefont {D.~H.}\ \bibnamefont {Kim}},\ }\bibfield  {title} {\bibinfo
  {title} {Mid-infrared properties of a ${\mathrm{vo}}_{2}$ film near the
  metal-insulator transition},\ }\href
  {https://doi.org/10.1103/PhysRevB.54.4621} {\bibfield  {journal} {\bibinfo
  {journal} {Phys. Rev. B}\ }\textbf {\bibinfo {volume} {54}},\ \bibinfo
  {pages} {4621} (\bibinfo {year} {1996})}\BibitemShut {NoStop}%
\bibitem [{\citenamefont {Miura}\ \emph {et~al.}(2004)\citenamefont {Miura},
  \citenamefont {Osada},\ and\ \citenamefont {Takeyama}}]{Miura2003}%
  \BibitemOpen
  \bibfield  {author} {\bibinfo {author} {\bibfnamefont {N.}~\bibnamefont
  {Miura}}, \bibinfo {author} {\bibfnamefont {T.}~\bibnamefont {Osada}},\ and\
  \bibinfo {author} {\bibfnamefont {S.}~\bibnamefont {Takeyama}},\ }\bibfield
  {title} {\bibinfo {title} {Research in super-high pulsed magnetic fields at
  the megagauss laboratory of the university of tokyo},\ }\href
  {https://doi.org/https://doi.org/10.1023/A:1025689218138} {\bibfield
  {journal} {\bibinfo  {journal} {J. Low Temp. Phys}\ }\textbf {\bibinfo
  {volume} {133}},\ \bibinfo {pages} {139} (\bibinfo {year}
  {2004})}\BibitemShut {NoStop}%
\bibitem [{\citenamefont {Date}(1995)}]{Date1995}%
  \BibitemOpen
  \bibfield  {author} {\bibinfo {author} {\bibfnamefont {M.}~\bibnamefont
  {Date}},\ }\bibfield  {title} {\bibinfo {title} {Recent progress in high
  field magnetism},\ }\href {https://doi.org/https://doi.org/10.1071/PH950187}
  {\bibfield  {journal} {\bibinfo  {journal} {Australian Journal of Physics}\
  }\textbf {\bibinfo {volume} {48}},\ \bibinfo {pages} {187} (\bibinfo {year}
  {1995})}\BibitemShut {NoStop}%
\bibitem [{\citenamefont {Detmer}\ \emph {et~al.}(1997)\citenamefont {Detmer},
  \citenamefont {Schmelcher}, \citenamefont {Diakonos},\ and\ \citenamefont
  {Cederbaum}}]{Detmer1997}%
  \BibitemOpen
  \bibfield  {author} {\bibinfo {author} {\bibfnamefont {T.}~\bibnamefont
  {Detmer}}, \bibinfo {author} {\bibfnamefont {P.}~\bibnamefont {Schmelcher}},
  \bibinfo {author} {\bibfnamefont {F.~K.}\ \bibnamefont {Diakonos}},\ and\
  \bibinfo {author} {\bibfnamefont {L.~S.}\ \bibnamefont {Cederbaum}},\
  }\bibfield  {title} {\bibinfo {title} {Hydrogen molecule in magnetic fields:
  The ground states of the \ensuremath{\Sigma} manifold of the parallel
  configuration},\ }\href {https://doi.org/10.1103/PhysRevA.56.1825} {\bibfield
   {journal} {\bibinfo  {journal} {Phys. Rev. A}\ }\textbf {\bibinfo {volume}
  {56}},\ \bibinfo {pages} {1825} (\bibinfo {year} {1997})}\BibitemShut
  {NoStop}%
\bibitem [{\citenamefont {Eyert}\ \emph {et~al.}(2005)\citenamefont {Eyert},
  \citenamefont {Schwingenschl{\''{o}}gl},\ and\ \citenamefont
  {Eckern}}]{Eyert2005}%
  \BibitemOpen
  \bibfield  {author} {\bibinfo {author} {\bibfnamefont {V.}~\bibnamefont
  {Eyert}}, \bibinfo {author} {\bibfnamefont {U.}~\bibnamefont
  {Schwingenschl{\''{o}}gl}},\ and\ \bibinfo {author} {\bibfnamefont
  {U.}~\bibnamefont {Eckern}},\ }\bibfield  {title} {\bibinfo {title} {Covalent
  bonding and hybridization effect in the corundum-type transition-metal oxides
  v2o3 and ti2o3},\ }\href {https://doi.org/10.1209/epl/i2005-10050-2}
  {\bibfield  {journal} {\bibinfo  {journal} {Europhys. Lett.}\ }\textbf
  {\bibinfo {volume} {70}},\ \bibinfo {pages} {782 } (\bibinfo {year}
  {2005})}\BibitemShut {NoStop}%
\bibitem [{\citenamefont {Chang}\ \emph {et~al.}(2018)\citenamefont {Chang},
  \citenamefont {Koethe}, \citenamefont {Hu}, \citenamefont {Weinen},
  \citenamefont {Agrestini}, \citenamefont {Zhao}, \citenamefont {Gegner},
  \citenamefont {Ott}, \citenamefont {Panaccione}, \citenamefont {Wu},
  \citenamefont {Haverkort}, \citenamefont {Roth}, \citenamefont {Komarek},
  \citenamefont {Offi}, \citenamefont {Monaco}, \citenamefont {Liao},
  \citenamefont {Tsuei}, \citenamefont {Lin}, \citenamefont {Chen},
  \citenamefont {Tanaka},\ and\ \citenamefont {Tjeng}}]{Chang2018}%
  \BibitemOpen
  \bibfield  {author} {\bibinfo {author} {\bibfnamefont {C.~F.}\ \bibnamefont
  {Chang}}, \bibinfo {author} {\bibfnamefont {T.~C.}\ \bibnamefont {Koethe}},
  \bibinfo {author} {\bibfnamefont {Z.}~\bibnamefont {Hu}}, \bibinfo {author}
  {\bibfnamefont {J.}~\bibnamefont {Weinen}}, \bibinfo {author} {\bibfnamefont
  {S.}~\bibnamefont {Agrestini}}, \bibinfo {author} {\bibfnamefont
  {L.}~\bibnamefont {Zhao}}, \bibinfo {author} {\bibfnamefont {J.}~\bibnamefont
  {Gegner}}, \bibinfo {author} {\bibfnamefont {H.}~\bibnamefont {Ott}},
  \bibinfo {author} {\bibfnamefont {G.}~\bibnamefont {Panaccione}}, \bibinfo
  {author} {\bibfnamefont {H.}~\bibnamefont {Wu}}, \bibinfo {author}
  {\bibfnamefont {M.~W.}\ \bibnamefont {Haverkort}}, \bibinfo {author}
  {\bibfnamefont {H.}~\bibnamefont {Roth}}, \bibinfo {author} {\bibfnamefont
  {A.~C.}\ \bibnamefont {Komarek}}, \bibinfo {author} {\bibfnamefont
  {F.}~\bibnamefont {Offi}}, \bibinfo {author} {\bibfnamefont {G.}~\bibnamefont
  {Monaco}}, \bibinfo {author} {\bibfnamefont {Y.-F.}\ \bibnamefont {Liao}},
  \bibinfo {author} {\bibfnamefont {K.-D.}\ \bibnamefont {Tsuei}}, \bibinfo
  {author} {\bibfnamefont {H.-J.}\ \bibnamefont {Lin}}, \bibinfo {author}
  {\bibfnamefont {C.~T.}\ \bibnamefont {Chen}}, \bibinfo {author}
  {\bibfnamefont {A.}~\bibnamefont {Tanaka}},\ and\ \bibinfo {author}
  {\bibfnamefont {L.~H.}\ \bibnamefont {Tjeng}},\ }\bibfield  {title} {\bibinfo
  {title} {$c$-axis dimer and its electronic breakup: The insulator-to-metal
  transition in ${\mathrm{ti}}_{2}{\mathrm{o}}_{3}$},\ }\href
  {https://doi.org/10.1103/PhysRevX.8.021004} {\bibfield  {journal} {\bibinfo
  {journal} {Phys. Rev. X}\ }\textbf {\bibinfo {volume} {8}},\ \bibinfo {pages}
  {021004} (\bibinfo {year} {2018})}\BibitemShut {NoStop}%
\bibitem [{\citenamefont {Horibe}\ \emph {et~al.}(2006)\citenamefont {Horibe},
  \citenamefont {Shingu}, \citenamefont {Kurushima}, \citenamefont {Ishibashi},
  \citenamefont {Ikeda}, \citenamefont {Kato}, \citenamefont {Motome},
  \citenamefont {Furukawa}, \citenamefont {Mori},\ and\ \citenamefont
  {Katsufuji}}]{Horibe2006}%
  \BibitemOpen
  \bibfield  {author} {\bibinfo {author} {\bibfnamefont {Y.}~\bibnamefont
  {Horibe}}, \bibinfo {author} {\bibfnamefont {M.}~\bibnamefont {Shingu}},
  \bibinfo {author} {\bibfnamefont {K.}~\bibnamefont {Kurushima}}, \bibinfo
  {author} {\bibfnamefont {H.}~\bibnamefont {Ishibashi}}, \bibinfo {author}
  {\bibfnamefont {N.}~\bibnamefont {Ikeda}}, \bibinfo {author} {\bibfnamefont
  {K.}~\bibnamefont {Kato}}, \bibinfo {author} {\bibfnamefont {Y.}~\bibnamefont
  {Motome}}, \bibinfo {author} {\bibfnamefont {N.}~\bibnamefont {Furukawa}},
  \bibinfo {author} {\bibfnamefont {S.}~\bibnamefont {Mori}},\ and\ \bibinfo
  {author} {\bibfnamefont {T.}~\bibnamefont {Katsufuji}},\ }\bibfield  {title}
  {\bibinfo {title} {Spontaneous formation of vanadium ``molecules'' in a
  geometrically frustrated crystal: ${\mathrm{alv}}_{2}{\mathrm{o}}_{4}$},\
  }\href {https://doi.org/10.1103/PhysRevLett.96.086406} {\bibfield  {journal}
  {\bibinfo  {journal} {Phys. Rev. Lett.}\ }\textbf {\bibinfo {volume} {96}},\
  \bibinfo {pages} {086406} (\bibinfo {year} {2006})}\BibitemShut {NoStop}%
\bibitem [{\citenamefont {Radaelli}\ \emph {et~al.}(2002)\citenamefont
  {Radaelli}, \citenamefont {Horibe}, \citenamefont {Gutmann}, \citenamefont
  {Ishibashi}, \citenamefont {Chen}, \citenamefont {Ibberson}, \citenamefont
  {Koyama}, \citenamefont {Hor}, \citenamefont {Kiryukhin},\ and\ \citenamefont
  {Cheong}}]{Radaelli2002}%
  \BibitemOpen
  \bibfield  {author} {\bibinfo {author} {\bibfnamefont {P.~G.}\ \bibnamefont
  {Radaelli}}, \bibinfo {author} {\bibfnamefont {Y.}~\bibnamefont {Horibe}},
  \bibinfo {author} {\bibfnamefont {M.~J.}\ \bibnamefont {Gutmann}}, \bibinfo
  {author} {\bibfnamefont {H.}~\bibnamefont {Ishibashi}}, \bibinfo {author}
  {\bibfnamefont {C.~H.}\ \bibnamefont {Chen}}, \bibinfo {author}
  {\bibfnamefont {R.~M.}\ \bibnamefont {Ibberson}}, \bibinfo {author}
  {\bibfnamefont {Y.}~\bibnamefont {Koyama}}, \bibinfo {author} {\bibfnamefont
  {Y.-S.}\ \bibnamefont {Hor}}, \bibinfo {author} {\bibfnamefont
  {V.}~\bibnamefont {Kiryukhin}},\ and\ \bibinfo {author} {\bibfnamefont
  {S.-W.}\ \bibnamefont {Cheong}},\ }\bibfield  {title} {\bibinfo {title}
  {Formation of isomorphic ir3+ and ir4+ octamers and spin dimerization in the
  spinel cuir2s4},\ }\href {https://doi.org/10.1038/416155a} {\bibfield
  {journal} {\bibinfo  {journal} {Nature}\ }\textbf {\bibinfo {volume} {416}},\
  \bibinfo {pages} {155 } (\bibinfo {year} {2002})}\BibitemShut {NoStop}%
\bibitem [{\citenamefont {Muraoka}\ and\ \citenamefont
  {Hiroi}(2002)}]{Muraoka2002}%
  \BibitemOpen
  \bibfield  {author} {\bibinfo {author} {\bibfnamefont {Y.}~\bibnamefont
  {Muraoka}}\ and\ \bibinfo {author} {\bibfnamefont {Z.}~\bibnamefont
  {Hiroi}},\ }\bibfield  {title} {\bibinfo {title} {Metal-insulator transition
  of ${\mathrm{v}\mathrm{o}}_{2}$ thin films grown on
  ${\mathrm{ti}\mathrm{o}}_{2}$ (001) and (110) substrates},\ }\href@noop {}
  {\bibfield  {journal} {\bibinfo  {journal} {Applied Physics Letters}\
  }\textbf {\bibinfo {volume} {80}},\ \bibinfo {pages} {583} (\bibinfo {year}
  {2002})}\BibitemShut {NoStop}%
\bibitem [{\citenamefont {Hamaoui}\ \emph {et~al.}(2019)\citenamefont
  {Hamaoui}, \citenamefont {Horny}, \citenamefont {Gomez-Heredia},
  \citenamefont {Ramirez-Rincon}, \citenamefont {Ordonez-Miranda},
  \citenamefont {Champeaux}, \citenamefont {Dumas-Bouchiat}, \citenamefont
  {Alvarado-Gil}, \citenamefont {Ezzahri}, \citenamefont {K.},\ and\
  \citenamefont {Chirtoc}}]{Hamaoui2019}%
  \BibitemOpen
  \bibfield  {author} {\bibinfo {author} {\bibfnamefont {G.}~\bibnamefont
  {Hamaoui}}, \bibinfo {author} {\bibfnamefont {N.}~\bibnamefont {Horny}},
  \bibinfo {author} {\bibfnamefont {C.}~\bibnamefont {Gomez-Heredia}}, \bibinfo
  {author} {\bibfnamefont {J.}~\bibnamefont {Ramirez-Rincon}}, \bibinfo
  {author} {\bibfnamefont {J.}~\bibnamefont {Ordonez-Miranda}}, \bibinfo
  {author} {\bibfnamefont {C.}~\bibnamefont {Champeaux}}, \bibinfo {author}
  {\bibfnamefont {F.}~\bibnamefont {Dumas-Bouchiat}}, \bibinfo {author}
  {\bibfnamefont {J.}~\bibnamefont {Alvarado-Gil}}, \bibinfo {author}
  {\bibfnamefont {Y.}~\bibnamefont {Ezzahri}}, \bibinfo {author} {\bibfnamefont
  {J.}~\bibnamefont {K.}},\ and\ \bibinfo {author} {\bibfnamefont
  {M.}~\bibnamefont {Chirtoc}},\ }\bibfield  {title} {\bibinfo {title}
  {Thermophysical characterisation of v${\mathrm{o}}_{2}$ thin films hysteresis
  and its application in thermal rectification},\ }\href
  {https://doi.org/doi:10.1038/s41598-019-45436-0} {\bibfield  {journal}
  {\bibinfo  {journal} {Sci. Rep.}\ }\textbf {\bibinfo {volume} {9}},\ \bibinfo
  {pages} {8728 1} (\bibinfo {year} {2019})}\BibitemShut {NoStop}%
\bibitem [{\citenamefont {Brito}\ \emph {et~al.}(2016)\citenamefont {Brito},
  \citenamefont {Aguiar}, \citenamefont {Haule},\ and\ \citenamefont
  {Kotliar}}]{Brito2016}%
  \BibitemOpen
  \bibfield  {author} {\bibinfo {author} {\bibfnamefont {W.~H.}\ \bibnamefont
  {Brito}}, \bibinfo {author} {\bibfnamefont {M.~C.~O.}\ \bibnamefont
  {Aguiar}}, \bibinfo {author} {\bibfnamefont {K.}~\bibnamefont {Haule}},\ and\
  \bibinfo {author} {\bibfnamefont {G.}~\bibnamefont {Kotliar}},\ }\bibfield
  {title} {\bibinfo {title} {Metal-insulator transition in ${\mathrm{vo}}_{2}$:
  A $\mathrm{DFT}+\mathrm{DMFT}$ perspective},\ }\href
  {https://doi.org/10.1103/PhysRevLett.117.056402} {\bibfield  {journal}
  {\bibinfo  {journal} {Phys. Rev. Lett.}\ }\textbf {\bibinfo {volume} {117}},\
  \bibinfo {pages} {056402} (\bibinfo {year} {2016})}\BibitemShut {NoStop}%
\bibitem [{\citenamefont {Muraoka}\ \emph {et~al.}(2018)\citenamefont
  {Muraoka}, \citenamefont {Nagao}, \citenamefont {Yao}, \citenamefont
  {Wakita}, \citenamefont {Terashima}, \citenamefont {Yokoya}, \citenamefont
  {Kumigashira},\ and\ \citenamefont {Oshima}}]{Muraoka2018}%
  \BibitemOpen
  \bibfield  {author} {\bibinfo {author} {\bibfnamefont {Y.}~\bibnamefont
  {Muraoka}}, \bibinfo {author} {\bibfnamefont {H.}~\bibnamefont {Nagao}},
  \bibinfo {author} {\bibfnamefont {Y.}~\bibnamefont {Yao}}, \bibinfo {author}
  {\bibfnamefont {T.}~\bibnamefont {Wakita}}, \bibinfo {author} {\bibfnamefont
  {K.}~\bibnamefont {Terashima}}, \bibinfo {author} {\bibfnamefont
  {T.}~\bibnamefont {Yokoya}}, \bibinfo {author} {\bibfnamefont
  {H.}~\bibnamefont {Kumigashira}},\ and\ \bibinfo {author} {\bibfnamefont
  {M.}~\bibnamefont {Oshima}},\ }\bibfield  {title} {\bibinfo {title} {Fermi
  surface topology in a metallic phase of v${\mathrm{o}}_{2}$ thin films grown
  on tio2(001) substrates},\ }\href
  {https://doi.org/10.1038/s41598-018-36281-8} {\bibfield  {journal} {\bibinfo
  {journal} {Sci. Rep.}\ }\textbf {\bibinfo {volume} {8}},\ \bibinfo {pages}
  {17906} (\bibinfo {year} {2018})}\BibitemShut {NoStop}%
\bibitem [{\citenamefont {Verleur}\ \emph {et~al.}(1968)\citenamefont
  {Verleur}, \citenamefont {Barker},\ and\ \citenamefont
  {Berglund}}]{Verleur1968}%
  \BibitemOpen
  \bibfield  {author} {\bibinfo {author} {\bibfnamefont {H.~W.}\ \bibnamefont
  {Verleur}}, \bibinfo {author} {\bibfnamefont {A.~S.}\ \bibnamefont
  {Barker}},\ and\ \bibinfo {author} {\bibfnamefont {C.~N.}\ \bibnamefont
  {Berglund}},\ }\bibfield  {title} {\bibinfo {title} {Optical properties of
  v${\mathrm{o}}_{2}$ between 0.25 and 5 ev},\ }\href
  {https://doi.org/10.1103/PhysRev.172.788} {\bibfield  {journal} {\bibinfo
  {journal} {Phys. Rev.}\ }\textbf {\bibinfo {volume} {172}},\ \bibinfo {pages}
  {788} (\bibinfo {year} {1968})}\BibitemShut {NoStop}%
\end{thebibliography}%

\end{document}